\documentclass{elsart}
\usepackage{epsfig,psfig}

\def\gsim{\;\raise0.3ex\hbox{$>$\kern-0.75em\raise-1.1ex\hbox{$\sim$}}\;}
\def\lsim{\;\raise0.3ex\hbox{$<$\kern-0.75em\raise-1.1ex\hbox{$\sim$}}\;}

\begin{document}

\begin{frontmatter}
\hfill{MPI-PhT/2001-52, SISSA 9/2002/EP}

\title{Cosmological bounds on neutrino degeneracy improved by flavor
oscillations}
\author[Ferrara]{A.D.~Dolgov\thanksref{ITEP}},
\author[Oxford]{S.H.~Hansen},
\author[MPI]{S.~Pastor},
\author[Trieste]{S.T.~Petcov\thanksref{Sofia}},
\author[MPI]{G.G.~Raffelt},
\author[MPI]{D.V.~Semikoz\thanksref{INR}}
\address[Ferrara]{INFN section of Ferrara, Via del Paradiso 12, 44100
Ferrara, Italy} 
\address[Oxford]{NAPL, University of Oxford, Keble road, OX1 3RH,
Oxford, UK}
\address[MPI]{Max-Planck-Institut f\"ur Physik, F\"ohringer Ring 6,
80805 M\"unchen, Germany}
\address[Trieste]{Scuola Internazionale Superiore di Studi Avanzati
and INFN section of Trieste, Via Beirut 2--4, 34014 Trieste, Italy}
\thanks[ITEP]{Also at: ITEP, Bol.~Cheremushkinskaya 25, Moscow 117259,
Russia.}
\thanks[Sofia]{Also at: Institute of Nuclear Research and Nuclear
Energy, Bulgarian Academy of Sciences, 1784 Sofia, Bulgaria.}
\thanks[INR]{Also at: Institute for Nuclear Research of the Academy of
Sciences of Russia, 60th October Anniversary Prospect 7a, Moscow
117312, Russia.}

\begin{abstract}

We study three-flavor neutrino oscillations in the early universe in
the presence of neutrino chemical potentials. We take into account all
effects from the background medium, i.e.~collisional damping, the
refractive effects from charged leptons, and in particular neutrino
self-interactions that synchronize the neutrino oscillations.  We find
that effective flavor equilibrium between all active neutrino species
is established well before the big-bang nucleosynthesis (BBN) epoch if
the neutrino oscillation parameters are in the range indicated by the
atmospheric neutrino data and by the large mixing angle (LMA) MSW
solution of the solar neutrino problem. For the other solutions of the
solar neutrino problem, partial flavor equilibrium may be achieved if
the angle $\theta_{13}$ is close to the experimental limit
$\tan^2\theta_{13}\lsim 0.065$. In the LMA case, the BBN limit on the
$\nu_e$ degeneracy parameter, $|\xi_\nu|\lsim 0.07$, now
applies to all flavors. Therefore, a putative extra cosmic radiation
contribution from degenerate neutrinos is limited to such low values
that it is neither observable in the large-scale structure of the
universe nor in the anisotropies of the cosmic microwave background
radiation. Existing limits and possible future measurements, for
example in KATRIN, of the absolute neutrino mass scale will provide
unambiguous information on the cosmic neutrino mass density,
essentially free of the uncertainty of the neutrino chemical
potentials.
\end{abstract}
\end{frontmatter}


\section{Introduction}

\label{sec:introduction}

The cosmic matter and radiation inventory is known with ever
increasing precision, but many important questions remain open.  The
cosmic neutrino background is a generic feature of the standard hot
big bang model, and its presence is indirectly established by the
accurate agreement between the calculated and observed primordial
light-element abundances. However, the exact neutrino number density
is not known as it depends on the unknown chemical potentials for the
three flavors. (In addition there could be a population of sterile
neutrinos, a hypothesis that we will not discuss here.)  The standard
assumption is that the asymmetry between neutrinos and anti-neutrinos
is of order the baryon asymmetry $\eta \equiv
(n_B-n_{\bar{B}})/n_\gamma \simeq 6\times 10^{-10}$. This would be the
case, for example, if $B-L=0$ where $B$ and $L$ are the cosmic baryon
and lepton asymmetries, respectively. While $B-L=0$ is motivated by
scenarios where the baryon asymmetry is obtained via
leptogenesis~\cite{Buchmuller:2000as}, there are models for producing
large $L$ and small $B$
\cite{Harvey:1981cu,DK,Casas:1999gx,Dolgov:1991fr}.

In order to quantify a putative neutrino asymmetry we assume that well
before thermal neutrino decoupling a given flavor is characterized by
a Fermi-Dirac distribution with a chemical potential $\mu_\nu$,
$f_\nu(p,T) = \left [\exp \left(E_p/T -\xi_\nu \right)+1
\right]^{-1}$, where $\xi_\nu \equiv \mu_\nu/T$ is the degeneracy
parameter and $E_p\simeq p$ since we may neglect small neutrino mass
effects on the distribution function. For anti-neutrinos the
distribution function is given by $\xi_{\bar{\nu}}=-\xi_\nu$.

A neutrino chemical potential modifies the outcome of primordial
nucleosynthesis in two different ways~\cite{Sarkar:1996dd}.  The first
effect appears only in the electron sector because electron neutrinos
participate in the beta processes which determine the primordial
neutron-to-proton ratio so that $n/p\propto \exp(-\xi_e)$.\footnote{We
use the notation $\xi_e\equiv \xi_{\nu_e}$ etc.\ to avoid double
subscripts. We never discuss charged-lepton chemical potentials so
that there should be no confusion.}  Therefore, a positive $\xi_e$
decreases $Y_{\rm p}$, the primordial $^4$He mass fraction, while a
negative $\xi_e$ increases it, leading to an allowed range
\begin{equation}
-0.01<\xi_e<0.07~,
\label{stronglimit}
\end{equation}
compatible with $\xi_e=0$ (see
Refs.~\cite{Kang:1992xa,Esposito:2000hh,Esposito:2001sv,DiBari:2001ua}
and Sec.~\ref{sec:newlimits}).  A second effect is an increase of the
neutrino energy density for any non-zero $\xi$ which in turn increases
the expansion rate of the universe, thus enhancing $Y_{\rm p}$. This
applies to all flavors so that the effect of chemical potentials in
the $\nu_\mu$ or $\nu_\tau$ sector can be compensated by a positive
$\xi_e$. Altogether the big-bang nucleosynthesis (BBN) limits on the
neutrino chemical potentials are thus not very restrictive.

Another consequence of the extra radiation density in degenerate
neutrinos is that it postpones the epoch of matter-radiation equality.
In the cosmic microwave background radiation (CMBR) it boosts the
amplitude of the first acoustic peak of the angular power spectrum and
shifts all peaks to smaller scales. Moreover, the power spectrum of
density fluctuations on small scales is
suppressed~\cite{Lesgourgues:1999wu,Hu:1999tk}, 
leading to observable effects in
the cosmic large-scale structure~(LSS).

A recent analysis of the combined effect of a non-zero neutrino
asymmetry on BBN and CMBR/LSS yields the allowed
regions~\cite{Hansen:2001hi}
\begin{equation}
-0.01 < \xi_e < 0.22, \qquad
|\xi_{\mu,\tau}| < 2.6,
\label{nooscbounds}
\end{equation}
in agreement with similar bounds
in~\cite{Hannestad:2001hn,Kneller:2001cd}. These limits allow for a
very significant radiation contribution of degenerate neutrinos,
leading many authors to discuss the implications of a large neutrino
asymmetry in different physical situations. These include the
explanation of the former discrepancy between the BBN and CMBR results
on the baryon asymmetry \cite{Lesgourgues:2000eq} or the origin of the
cosmic rays with energies in excess of the Greisen-Zatsepin-Kuzmin
cutoff~\cite{Gelmini:1999qa}. In addition, if the present relic
neutrino background is strongly degenerate, it would enhance the
contribution of massive neutrinos to the total energy
density~\cite{Pal:1999jv,Lesgourgues:2000ej} and affect the flavor
oscillations of the high-energy neutrinos~\cite{Lunardini:2001fy}
which are thought to be produced in the astrophysical accelerators of
high-energy cosmic rays.

The limits in Eq.~(\ref{nooscbounds}) ignore neutrino flavor
oscillations, an assumption which is no longer justified in view of
the experimental signatures for neutrino oscillations by solar and
atmospheric neutrinos. For zero initial neutrino chemical potentials,
the flavor neutrinos have the same spectra so that oscillations
produce no effect. This is true up to a small spectral distortion
caused by the heating of neutrinos from $e^+e^-$ annihilations, an
effect which is different for electron and muon/tau neutrinos and
which causes a small relative change in the final production of $^4$He
of order $10^{-3}$~\cite{Dolgov:1997mb}. This relative change is
slightly enhanced by neutrino flavor
oscillations~\cite{Langacker:1987jv,Hannestad:2001iy}. In the
presence of neutrino asymmetries, flavor oscillations equalize the
neutrino chemical potentials if there is enough time for this
relaxation process to be effective~\cite{Savage:1991by}. If flavor
equilibrium is reached before BBN, then the restrictive limits on
$\xi_e$ in Eq.~(\ref{stronglimit}) will apply to all flavors, in turn
implying that the cosmic neutrino radiation density is close to its
standard value.  As a consequence, it is no longer necessary to use
the neutrino radiation density as a fit parameter for CMBR/LSS
analyses, unless one considers exotic models with decaying massive
particles.

The effects of flavor oscillations on possible neutrino degeneracies
have been considered in~\cite{Lunardini:2001fy}, where it was
concluded that flavor equilibrium was achieved before the BBN epoch if
the solar neutrino problem was explained by the large-mixing angle
(LMA) solution. The LMA solution is favored by the current solar
neutrino data. Thus, it was concluded that in the LMA case a large
cosmic neutrino degeneracy was no longer allowed.

We revisit this problem because the flavor evolution of the neutrino
ensemble is more subtle than previously envisaged if medium effects
are systematically included.  Contrary to the treatment of
Ref.~\cite{Lunardini:2001fy}, the refractive effect of charged leptons
can not be ignored, and actually is one of the dominant effects.
While the background neutrinos produce an even larger refractive term,
its effect is to synchronize the neutrino
oscillations~\cite{Samuel:1993uw,Pastor:2001iu} which remain sensitive
to the charged-lepton contribution. Still, equilibrium is essentially,
but not completely, achieved in the LMA case so that our final
conclusion is qualitatively similar to that of
Ref.~\cite{Lunardini:2001fy}.

One counter-intuitive subtlety is that the neutrino self-potential
actually can suppress oscillations in a situation where the excess of
neutrinos in one flavor is exactly matched by an excess of
anti-neutrinos in another flavor. In this case the synchronized
oscillation frequency is zero so that oscillations begin only once the
cosmic expansion has diluted the self-term. Therefore, one can
construct cases where flavor equilibrium is not achieved before BBN
even in the LMA case.  However, this is only possible for specially
chosen initial conditions where $|\xi_\nu|$ is equal for all flavors,
but the absolute signs may be different. For the purpose of deriving
limits on $|\xi_\nu|$, however, this case is equivalent to the one
where equilibrium is achieved, the only important point being that
$|\xi_\nu|$ is approximately equal for all flavors at the BBN epoch.

Another subtlety appears in a full three-flavor analysis. We show that
achieving equilibrium in the LMA case does not depend on the value of
the mixing angle $\theta_{13}$, to which strict limits from reactor
experiments apply. Moreover, for the non-LMA solutions of the solar
neutrino problem, partial flavor equilibrium may be reached if the
angle $\theta_{13}$ is small but non-zero.

In Sec.~\ref{sec:neutrinoflavor} we set up the formalism to study
primordial flavor oscillations. We then turn in
Sec.~\ref{sec:twoflavor} to the primordial flavor evolution of a
simplified system where $\nu_e$ mixes maximally with one other flavor.
This two-flavor case will illustrate many of the important subtleties
of our problem.  Then we turn in Sec.~\ref{sec:threeflavor} to
realistic three-flavor situations which involve yet further
complications.  In Sec.~\ref{sec:newlimits} we finally derive new
limits on the degeneracy parameters and summarize our findings.


\section{Neutrino flavor oscillations in the early universe}

\label{sec:neutrinoflavor}

In order to study neutrino oscillations in the early universe we
characterize the neutrino ensemble in the usual way by generalized
occupation numbers, i.e.\ by $3 {\times} 3$ density matrices for
neutrinos and anti-neutrinos as described in
\cite{Sigl:1993fn,McKellar:1994ja}.  The form of the density matrices
for a mode with momentum $p$ is
\begin{equation}
\rho(p,t) = \left (\matrix{\rho_{ee} 
& \rho_{e\mu}& \rho_{e\tau}\cr
\rho_{\mu e}& \rho_{\mu\mu}& \rho_{\mu\tau}\cr
\rho_{\tau e} &\rho_{\tau \mu}& \rho_{\tau \tau} }\right),
\qquad \bar\rho(p,t) = 
\left (\matrix{\bar\rho_{ee} 
& \bar\rho_{\mu e}& \bar\rho_{\tau e}\cr
\bar\rho_{e\mu}& \bar\rho_{\mu\mu}& \bar\rho_{\tau\mu}\cr
\bar\rho_{e\tau} &\bar\rho_{\mu\tau}& \bar\rho_{\tau \tau} }\right)
\label{3by3} 
\end{equation}
where overbarred quantities refer to anti-neutrinos.  The diagonal
elements of the density matrices correspond to the usual occupation
numbers of the different flavors. The definition of the density matrix
for anti-neutrinos is ``transposed'' relative to that for neutrinos,
allowing one to write the equations of motion in a compact
form~\cite{Sigl:1993fn}.

The equations of motion for the density matrices relevant for our
situation of interest are~\cite{Sigl:1993fn,Pantaleone:1992eq}
\begin{eqnarray}\label{eq:3by3evol}
i\partial_t\rho_p&=&
+\left[\frac{M^2}{2p},\rho_p\right]
+\sqrt2 G_{\rm F}\left[\left(
-\frac{8p}{3 m_{\rm W}^2}{E}+\rho-\bar\rho
\right),\rho_p\right]
+{C}[\rho_p]~,\nonumber\\
i\partial_t\bar\rho_p&=&
-\left[\frac{M^2}{2p},\bar\rho_p\right]
+\sqrt2 G_{\rm F}\left[\left(
+\frac{8p}{3 m_{\rm W}^2}{E}+\rho-\bar\rho
\right),\bar\rho_p\right]
+{C}[\bar\rho_p]~,
\end{eqnarray}
where we use the notation $\rho_p=\rho(p,t)$ and $[{\cdot},{\cdot}]$
is the usual commutator. Further, $M^2$ is the mass-squared matrix in
the flavor basis; in the mass basis it would be ${\rm
diag}(m_1^2,m_2^2,m_3^2)$.  The diagonal matrix $E$ represents the
energy densities of charged leptons. For example, $E_{ee}$ is the
energy density of electrons plus that of positrons. The density matrix
$\rho$ is the integrated neutrino density matrix so that, for example,
$\rho_{ee}$ is the $\nu_e$ number density while $\bar\rho_{ee}$ is the
$\bar\nu_e$ number density. The term proportional to $(\rho-\bar\rho)$
is non-linear in the neutrino density matrix and represents
self-interactions. Finally, $C[{\cdot}]$ is the collision term which
is proportional to $G_{\rm F}^2$. We have neglected a refractive term
proportional to the neutrino energy density which is much smaller than
the $(\rho-\bar\rho)$ term in our present situation where the neutrino
asymmetries are assumed to be large. On the other hand, we have
neglected the usual term which is proportional to the charged-lepton
asymmetries. This asymmetric term is always negligible; at early times
(high temperatures) it is negligible compared to the $E$ term, while
at temperatures near $n/p$ freeze out ($T\simeq 1~{\rm MeV}$) it is
negligible compared to the vacuum term $M^2/2p$ if the mass-squared
differences coincide with those characterizing the atmospheric
neutrino and the solar neutrino LMA oscillations.

In the expanding universe we need to substitute
$\partial_t\to\partial_t-Hp\partial_p$ with $H$ the cosmic expansion
parameter. We have taken this into account rewriting the equations in
terms of comoving variables, as listed in the Appendix. We solve these
equations numerically, calculating the evolution of $\rho$ and
$\bar\rho$ on a grid of neutrino momenta.


\section{Two-flavor oscillations}

\label{sec:twoflavor}

\subsection{Equations of motion for the polarization vectors}

In order to develop a first understanding of flavor oscillation in the
early universe it will be useful to study first a two-flavor situation
involving $\nu_e$ and $\nu_\mu$. The usual relation between flavor and
mass eigenstates is
\begin{eqnarray}
\nu_e&=&\cos \theta ~\nu_1 + \sin \theta ~\nu_2~,\nonumber\\
\nu_\mu &=& -\sin \theta ~\nu_1 + \cos \theta ~\nu_2~.
\label{2flavors}
\end{eqnarray}
The $2 {\times} 2$ density matrices are
\begin{eqnarray}
\rho(p,t) = \left (\matrix{\rho_{ee}& \rho_{e\mu}\cr
\rho_{\mu e}& \rho_{\mu \mu} }\right) =
\frac{1}{2} \left[P_0(p,t) +\hbox{\boldmath$\sigma$}
\cdot{\bf P}(p,t)\right]~,
\nonumber \\
\bar{\rho}(p,t) = \left (\matrix{\bar{\rho}_{ee}&\bar{\rho}_{\mu e}\cr
\bar{\rho}_{e\mu}& \bar{\rho}_{\mu\mu} }\right) =
\frac{1}{2} \left[\overline{P}_0(p,t) +\hbox{\boldmath$\sigma$}\cdot
\overline{\bf P}(p,t)\right]~.
\label{2by2mutau} 
\end{eqnarray}
Here, $\sigma_i$ are the Pauli matrices while ${\bf P}(p,t)$ and
$\overline{\bf P}(p,t)$ are the usual polarization vectors for the
neutrino and anti-neutrino modes $p$, respectively. We normalize
$\rho$ and $\bar{\rho}$ to a Fermi-Dirac distribution with zero
chemical potential, $f(p) = \left [\exp
\left(p/T\right)+1\right]^{-1}$, so that for instance
$f_{\nu_a}(p,t)=\rho_{aa}(p,t)f(p)$. For the polarization vectors, the
equations of motion are given by the usual spin-precession formula
\begin{eqnarray}
\partial_t {\bf P}_p&=&+
\left[\frac{\Delta m^2}{2p}\,{\bf B}-
\frac{8\sqrt2 G_{\rm F} p}{3 m_{\rm W}^2} E_{ee}\,\hat{\bf z}\right]
\times{\bf P}_p\nonumber\\
&&\kern12em{}+\sqrt2\,G_{\rm F}({\bf P}-\overline{\bf P})\times {\bf P}_p 
+ C[{\bf P}_p]~,\nonumber\\
\partial_t \overline{\bf P}_p&=&-
\left[\frac{\Delta m^2}{2p}\,{\bf B}-
\frac{8\sqrt2 G_{\rm F} p}{3 m_{\rm W}^2} E_{ee}\,\hat{\bf z}\right]
\times\overline{\bf P}_p\nonumber\\
&&\kern12em{}+\sqrt2\, G_{\rm F}
({\bf P}-\overline{\bf P})\times \overline{\bf P}_p + C[\overline{\bf P}_p]~,
\label{polvec}
\end{eqnarray}
where $\Delta m^2=m_2^2-m_1^2$ and ${\bf
B}=(\sin2\theta,0,-\cos2\theta)$ with $\theta$ the vacuum mixing angle.
Further, $\hat {\bf z}$ is a unit vector in the $z$-direction,
$E_{ee}$ the electron-positron energy density, and ${\bf P}$ and 
$\overline{\bf P}$ are the integrated polarization vectors.

We will see that a detailed treatment of the collision terms is not
crucial. Therefore, we approximate them with a simple damping prescription
of the form
\begin{equation}
C[{\bf P}_p] = -D_p {\bf P}_p^{\rm T} 
\end{equation}
where the transverse part ${\bf P}_p^{\rm T}$ consists of the
$x$-$y$-projection of ${\bf P}_p$.  We use the damping functions as,
for instance, in \cite{McKellar:1994ja,Bell:1999ds}, but with slightly
modified coefficients.  We neglect a small dependence on $\xi$,
i.e.~we use the same damping functions for neutrinos and
antineutrinos.  We have checked that our results are insensitive to
the inclusion of the repopulation function, i.e.~$\dot{P}_0$ and
$\dot{\overline{P}}_0$ as given for instance in
\cite{McKellar:1994ja,Bell:1999ds}. Moreover, we did not include the
neutrino heating from $e^+e^-$ annihilations, since it is small even
in the presence of degeneracies~\cite{Esposito:2000hi}.

\subsection{Evolution of simple flavor asymmetry}

\label{sec:simpleflavor}

As a first case we consider a simple situation where initially
electron neutrinos do not have a chemical potential ($\xi_e=0$) while
muon neutrinos are asymmetric ($\xi_\mu=-0.05$). We use a rather small
$\nu_\mu$ asymmetry so that the anticipated equilibrium state
$\xi_e=\xi_\mu$ will be close to the BBN limit on $\xi_e$ in
Eq.~(\ref{nooscbounds}).  The quantitative evolution with a large
$\xi_\mu$ is the same.  Moreover, we use maximal mixing and $\Delta
m^2=4.5\times10^{-5}$ eV$^2$ as suggested by the LMA solution of the
solar neutrino problem.

\begin{figure}[t]
\begin{center}
\epsfig{file=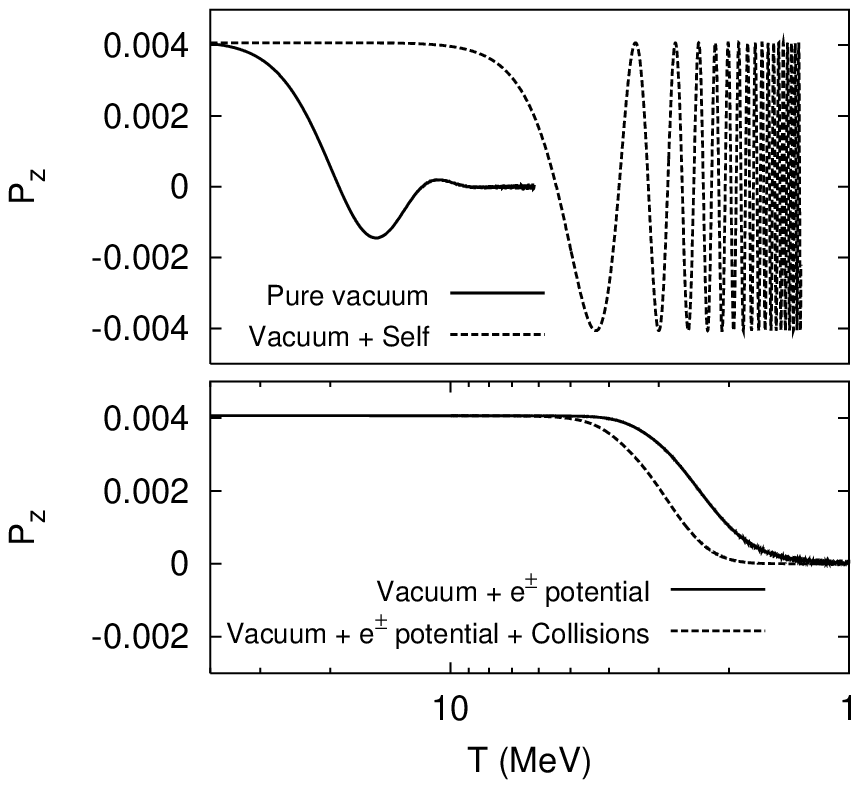,width=1.\textwidth}
\end{center}
\caption{Evolution of the $z$-component of the integrated polarization
vector ${\bf P}$ for different situations as explained in the text.}
\label{fig:1}
%
\bigskip
\epsfig{file=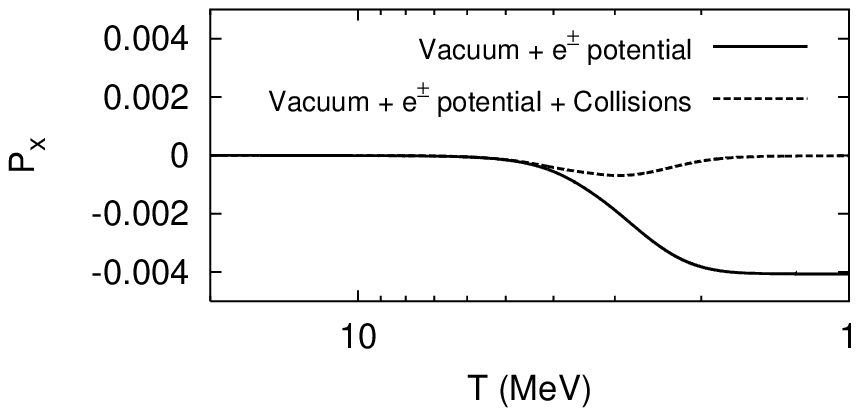,width=.7\textwidth}
\caption{Evolution of the $x$-component of the integrated polarization
vector ${\bf P}$ for the same cases as in the lower panel of 
Fig.~\ref{fig:1}.}
\label{fig:2}
\end{figure}

In order to illustrate the relative importance of the various
contributions to the equation of motion we first calculate the
primordial evolution of the integrated polarization vectors ${\bf P}$
and $\overline{\bf P}$ for pure vacuum oscillations, including only
the $\Delta m^2/2p$ terms in Eq.~(\ref{polvec}), without any medium
effects whatsoever. Our initial conditions $\xi_e=0$ and
$\xi_\mu=-0.05$ imply that there are more $\nu_e$ than $\nu_\mu$,
i.e.~${\bf P}$ points initially in the $+\hat{\bf z}$ direction,
while there are more $\bar\nu_\mu$ than $\bar\nu_e$, i.e.\
$\overline{\bf P}$ points initially in the $-\hat{\bf z}$ direction.
In the upper panel of Fig.~\ref{fig:1} we show the evolution of $P_z$
as a function of cosmic temperature; the evolution of $\overline{P}_z$
is similar, except with a negative initial value.  The oscillations
begin when the expansion rate has become slow enough at about $T\simeq
30$ MeV. The oscillation frequencies are different for different
modes, leading to quick decoherence and thus to an incoherent equal
mix of both flavors.  Evidently flavor equilibrium is reached long
before $n/p$ freeze-out at $T\simeq 1$ MeV.

As a next step we include the neutrino self-potential proportional to
${\bf P} - \overline{\bf P}$. The evolution is shown as a dotted line
in the upper panel of Fig.~\ref{fig:1}. It is now dominated by the
effect of synchronized oscillations, i.e.\ the self-potential forces
all neutrino modes to follow the same
oscillation~\cite{Pastor:2001iu}.  Because ${\bf P}$ and
$\overline{\bf P}$ point in opposite directions, the common or
synchronized oscillation frequency $\omega_{\rm synch}$ is very much
smaller than a typical $\Delta m^2/2p$. Explicitly we
have~\cite{Pastor:2001iu}
\begin{equation}
\omega_{\rm synch}=
\frac{1}{|{\bf I}|}
\int~\frac{d^3{p}}{(2\pi)^3}~f(p)
\frac{\Delta m^2}{2p}\,
{\bf\hat I}\cdot ({\bf P}_p+{\bf \overline P}_p)
\equiv \frac{\Delta m^2}{2p_{\rm eff}}
\label{wsynchantinu}
\end{equation}
where ${\bf I} \equiv {\bf P}-{\bf \overline P}$ and ${\bf\hat I}$ is
a unit vector in the direction of ${\bf I}$.  Numerically our initial
conditions imply $p_{\rm eff}\simeq 132\,T$.

Next we add to the vacuum term the effective potential caused by the
$e^\pm$ background [see Eq.~(\ref{eq:3by3evol})], without including
the self-term. The $e^\pm$ potential always points in the
$z$-direction, thereby suppressing flavor oscillations in the usual
way. This effect is shown in the lower panel of Fig.~\ref{fig:1}
(solid line).  ${\bf P}$~stays frozen at its initial value up to
$T\simeq 3$~MeV where the medium potential becomes unimportant
compared to the vacuum term.  At these temperatures collisions are
also about to become unimportant.  Therefore it is not surprising that
including collisions (dotted line) does not dramatically change the
evolution of~$P_z$.

However, this impression is misleading as can be seen from
Fig.~\ref{fig:2} where we show the evolution of $P_x$ for the same
cases as in the lower panel of Fig.~\ref{fig:1}. Without collisions
(solid line) the evolution of ${\bf P}$ is a simple turning from the
$z$- to the $x$-direction. The $e^\pm$ potential adiabatically
disappears, leading to the usual MSW-type evolution. In this case the
final result corresponds to equal numbers of both flavors, yet a
coherent flavor superposition. With collisional damping (dotted line)
the $x$-component is damped, leaving little flavor coherence of the
final state. Therefore, with or without collisional damping we reach
``flavor equilibrium'' in the sense of equal densities of both
flavors, but only collisions ensure the damping of the otherwise large
transverse part of the final ${\bf P}$.

\begin{figure}[t]
\epsfig{file=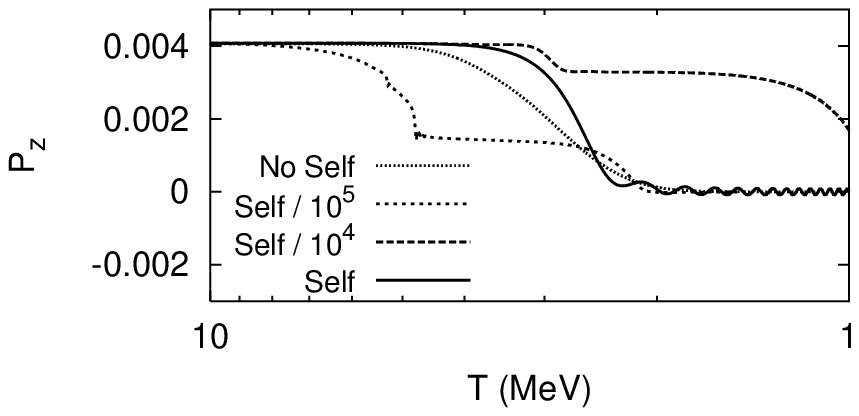,width=.7\textwidth}
\caption{Evolution of the $z$-component of the integrated polarization
vector ${\bf P}$ in the presence of the vacuum term, the $e^+e^-$
potential, collisional damping, and varying strength of the
neutrino self-interaction.}
\label{fig:3}
%
\bigskip
\epsfig{file=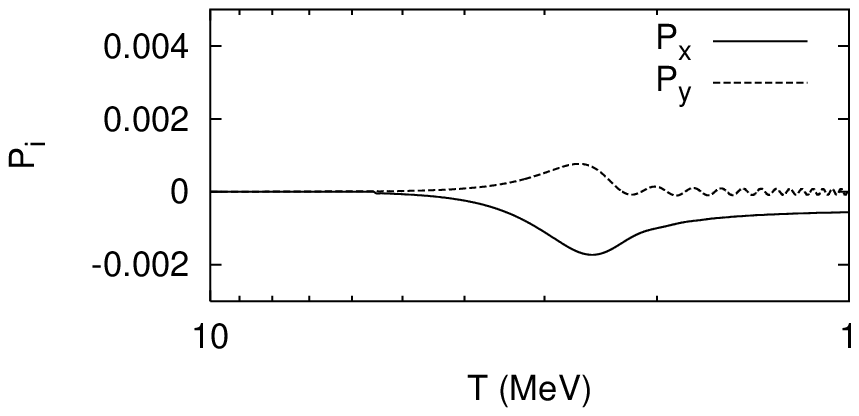,width=.7\textwidth}
\caption{Evolution of the $x$ and $y$ components of the integrated
polarization vector ${\bf P}$ in the presence of all effects and
full-strength neutrino self-interactions, corresponding to the
solid line in Fig.~\ref{fig:3}.}\label{fig:4}
\end{figure}

Finally we study how oscillations evolve in the presence of all
effects: vacuum, $e^\pm$ background, neutrino self-potential and
collisions. The results are shown in Fig.~\ref{fig:3} for different
strengths of the self-potential.  In one run it was switched off (``No
Self''), it was taken at full strength (``Self''), or it was
suppressed by a factor $10^4$ or $10^5$.  When the self-interaction
term is close to its real value (solid line), again all modes
oscillate in synch so that we see a combination of the synchronized
oscillations and the MSW-like effect of the background
medium. However, collisions ensure that no large transverse component
survives---see the corresponding evolution of $P_x$ and $P_y$ in
Fig.~\ref{fig:4}. Almost perfect flavor equilibrium is achieved with
or without self-interactions.

This conclusion is not to be taken for granted. When the
self-interaction term is artificially adjusted to be of comparable
strength to the other terms (vacuum or $e^\pm$ background), the
evolution shows complicated features (Fig.~\ref{fig:3}).  After some
initial conversion, $P_z$ remains constant for a long time, only to
reach equilibrium at a much later time. This behavior is not easy to
understand as it depends on all the ingredients of our calculation,
i.e.\ the background potential, the neutrino self-potential and
damping by collisions. We believe that the long phase where the
polarization vector essentially stands still is caused by the
synchronized oscillation frequency becoming very small. This is
strictly the case in a situation where the chemical potentials for two
flavors are equal but opposite---see Sec.~\ref{sec:equalopposite}
below. We believe that in the present case a combination of
decoherence of some of the modes and collisions drives the system into
a state where the synchronized oscillations frequency vanishes so that
oscillations stop until the cosmic expansion has diluted the neutrino
self-term enough to eliminate this effect.

\subsection{Equal but opposite asymmetries}

\label{sec:equalopposite}

Our discussion thus far suggests that for the parameters of interest
it is generic to achieve flavor equilibrium before the BBN epoch, even
though the loss of coherence may not always be complete.  However,
this conclusion depends on the simple initial conditions that we have
used thus far.  The different neutrino flavors may show large
asymmetries, yet the total cosmic lepton asymmetry could still be
small if initially $\xi_e=-\xi_\mu$, which corresponds to conservation
of the lepton number $L_e-L_\mu$. These initial conditions imply that
${\bf P}=-\overline{\bf P}$ so that the self-potential term $({\bf
P}-\overline{\bf P})$ is large while for each mode $({\bf
P}_p+\overline{\bf P}_p)=0$~\cite{Pastor:2001iu}. This implies, in
turn, that the synchronized oscillation frequency $\omega_{\rm
synch}=0$.  Therefore, in this case the synchronization effect caused
by the self-potential prevents flavor oscillations entirely, at least
until the self-term becomes weak, long after the BBN epoch.  While
flavor equilibrium is not achieved in this case, from the start we
have $|\xi_e|=|\xi_\mu|$ so that the BBN limits on $|\xi_e|$ would
apply to both flavors.


\section{Three-flavor oscillations}

\label{sec:threeflavor}

\subsection{Oscillation parameters}

We now turn to the more interesting case of three-flavor neutrino
oscillations. The neutrino flavor eigenstates $\nu_e$, $\nu_\mu$, and
$\nu_\tau$ are related to the mass eigenstates via the mixing matrix
\begin{equation}
\left(
    \begin{array}{ccc}                                 
        c_{12} c_{13}                
        & s_{12} c_{13} 
        & s_{13} \\
        -s_{12} c_{23} - c_{12} s_{23} s_{13} 
        & c_{12} c_{23} - s_{12} s_{23} s_{13}
        & s_{23} c_{13} \\
        s_{12} s_{23} - c_{12} c_{23} s_{13}
        & -c_{12} s_{23} - s_{12} c_{23} s_{13}
        & c_{23} c_{13}
\end{array}\right) \;.
\end{equation}
Here $c_{ij}=\cos \theta_{ij}$ and $s_{ij}=\sin \theta_{ij}$ for
$ij=12$, 23, or 13, and we have assumed CP conservation. The set of
oscillation parameters is now five-dimensional (see for
instance~\cite{Gonzalez-Garcia:2001sq}),
\begin{equation} \begin{array}{ll} 
   \label{oscpardef}
    \Delta m^2_{\rm sun} & \equiv \Delta m^2_{21} = m^2_2 - m^2_1 \\
    \Delta m^2_{\rm atm}   & \equiv \Delta m^2_{32} = m^2_3 - m^2_2 \\
    \theta_{\rm sun}     & \equiv \theta_{12} \\
    \theta_{\rm atm}       & \equiv \theta_{23} \\
    \theta_{13}
\end{array} 
\end{equation}
We do not perform a global analysis of all possible values of these
parameters, but fix them to be in the regions that solve the
atmospheric and solar neutrino
problems~\cite{Gonzalez-Garcia:2001sq,Fogli:2001xt}.  In particular we
take $\Delta m^2_{\rm atm}=3 \times 10^{-3}~\mbox{eV}^2$ and maximal
mixing for $\theta_{\rm atm}$ from the former, while from the solar
analyses we consider the following values for $\Delta m^2_{\rm sun}$
in eV$^2$: $4.5\times 10^{-5}$, $7 \times 10^{-6}$, $1 \times 10^{-7}$,
$8 \times 10^{-11}$ for the Large Mixing Angle (LMA), Small Mixing
Angle (SMA), LOW and Vacuum regions, respectively. For the angle
$\theta_{\rm sun}$ we take the approximation of maximal mixing for all
cases except SMA where we use $\theta_{\rm sun}=1.5^\circ$.

\subsection{Simple three flavor case}
\begin{figure}[t]
\centerline{\epsfig{file=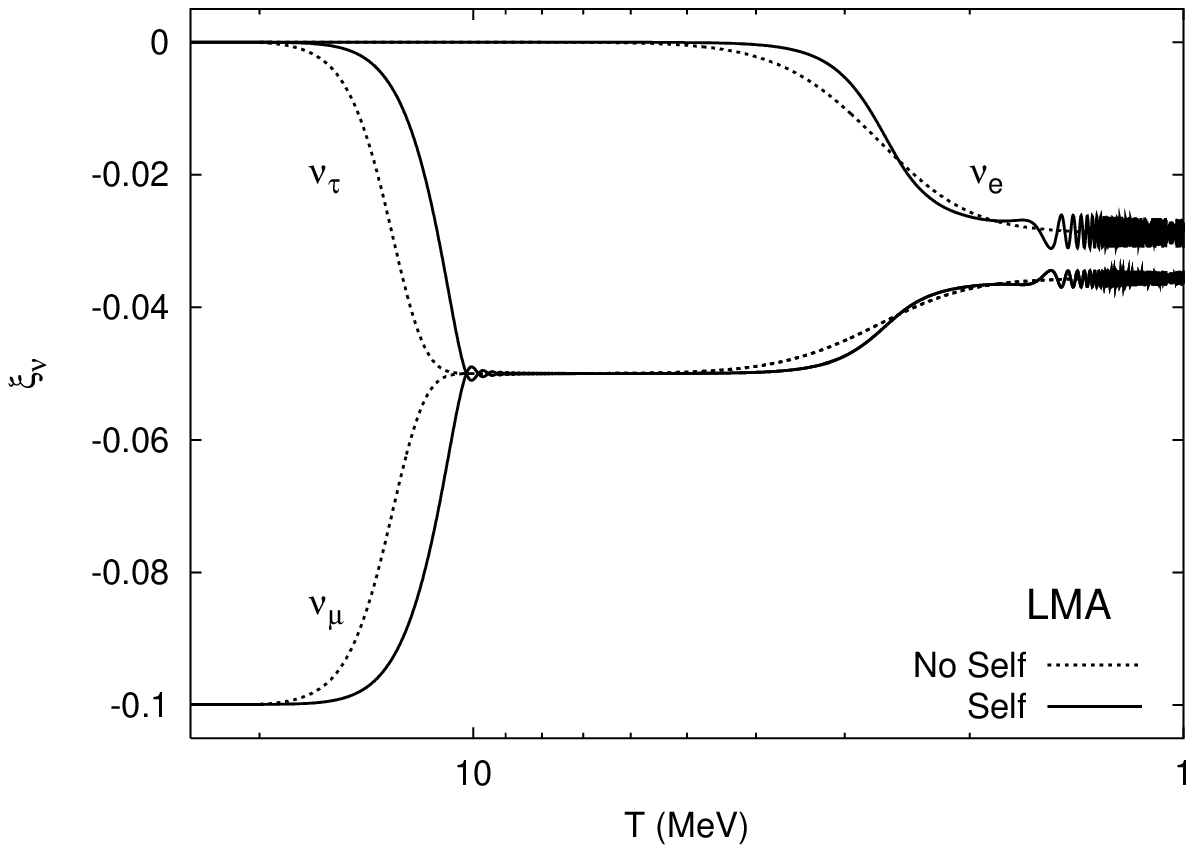,width=12cm}}
\caption{Evolution of the neutrino degeneracy parameters for the LMA
case, $\theta_{13}= 0$, and the initial values $\xi_e=\xi_\tau=0$ and
$\xi_\mu=-0.1$.}
\label{fig:5}
%
\bigskip
\centerline{\epsfig{file=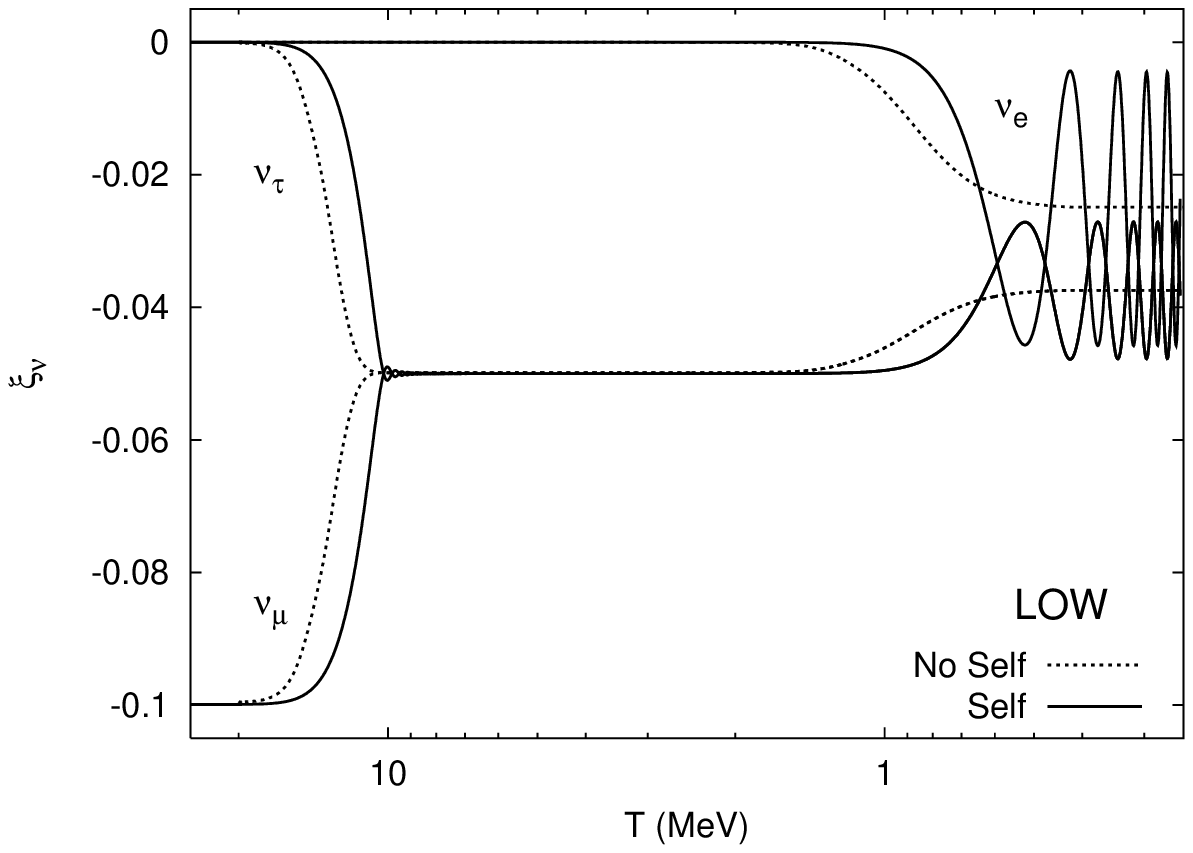,width=12cm}}
\caption{Evolution of the neutrino degeneracy parameters for the LOW
case, $\theta_{13}= 0$, and the initial values $\xi_e=\xi_\tau=0$ and
$\xi_\mu=-0.1$.}
\label{fig:6}
\end{figure}
We begin with the simplified situation where initially only the muon
neutrinos are asymmetric ($\xi_\mu=-0.1$). We perform a full
three-flavor calculation, but for now set $\theta_{13}=0$. The
evolution of the neutrino asymmetries is shown in Figs.~\ref{fig:5}
and~\ref{fig:6} for the LMA and LOW cases, respectively, both with and
without the neutrino self-interactions. For this choice of oscillation
parameters the three-flavor oscillations effectively separate as two
two-flavor problems for the atmospheric and solar parameters,
respectively. The oscillations caused by the largest $\Delta m^2$ are
effective at $T \simeq 20$ MeV, as soon as the $\mu^\pm$ background
disappears completely. The presence of the self-term causes only a
slight delay in the equilibration of $\xi_\mu$ and $\xi_\tau$.

The oscillations due to $\Delta m^2_{\rm sun}$ and $\theta_{\rm sun}$
are effective only when the vacuum term overcomes the $e^\pm$
potential. In the LMA case, the conversions takes place above $T
\simeq 1$~MeV, leading to nearly complete flavor equilibrium before
the onset of BBN. For the LOW parameters the synchronized
oscillations just start at that epoch. The presence of the neutrino
self-potential does not significantly change the picture in the LMA
case while for the LOW case one clearly observes the phenomenon of
synchronized oscillations.  For the SMA and Vacuum regions primordial
oscillations involving $\nu_e$ are not effective before BBN if
$\theta_{13}=0$.

\subsection{Non-zero 13 mixing}

\begin{figure}[t]
\centerline{\epsfig{file=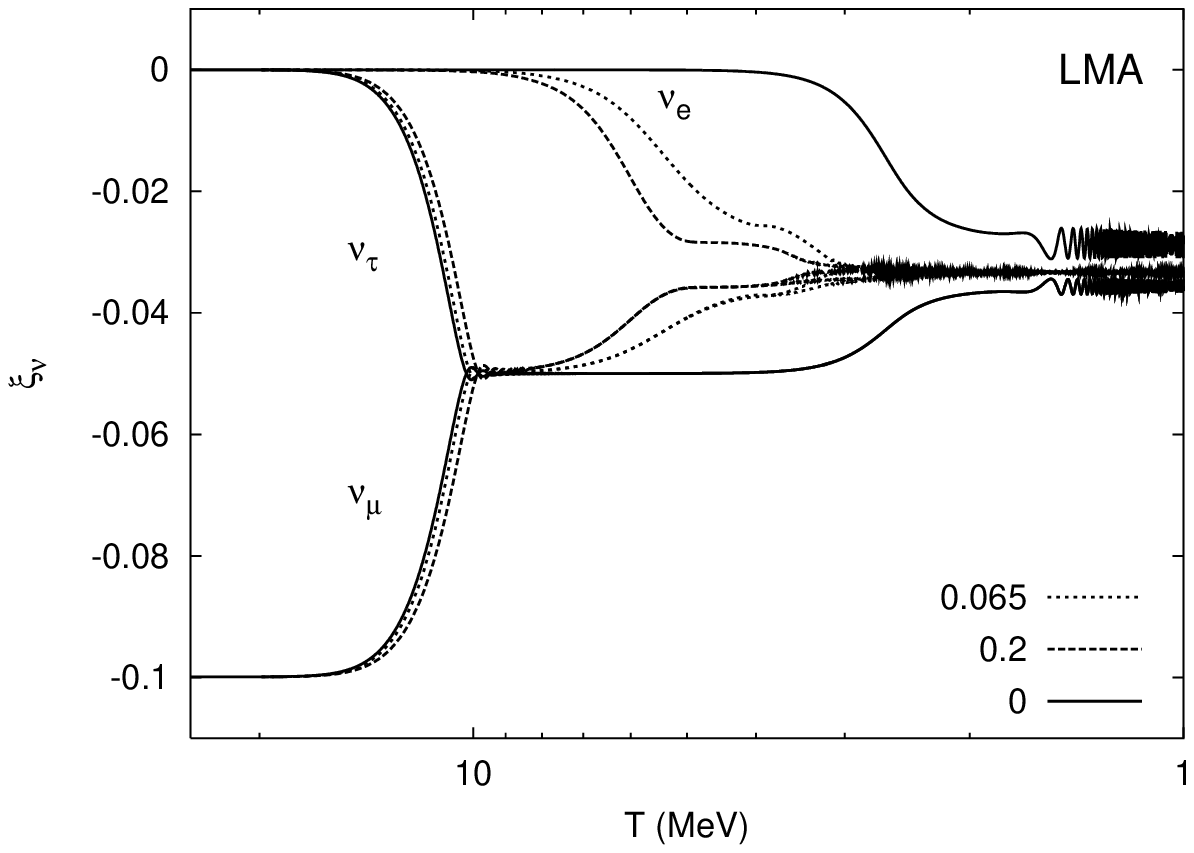,width=12cm}}
\caption{Evolution of the neutrino degeneracies for different values
of $\tan^2 \theta_{13}$ in the LMA case for the initial conditions 
$\xi_e=\xi_\tau=0$ and $\xi_\mu=-0.1$.}
\label{fig:7}
%
\bigskip
\centerline{\epsfig{file=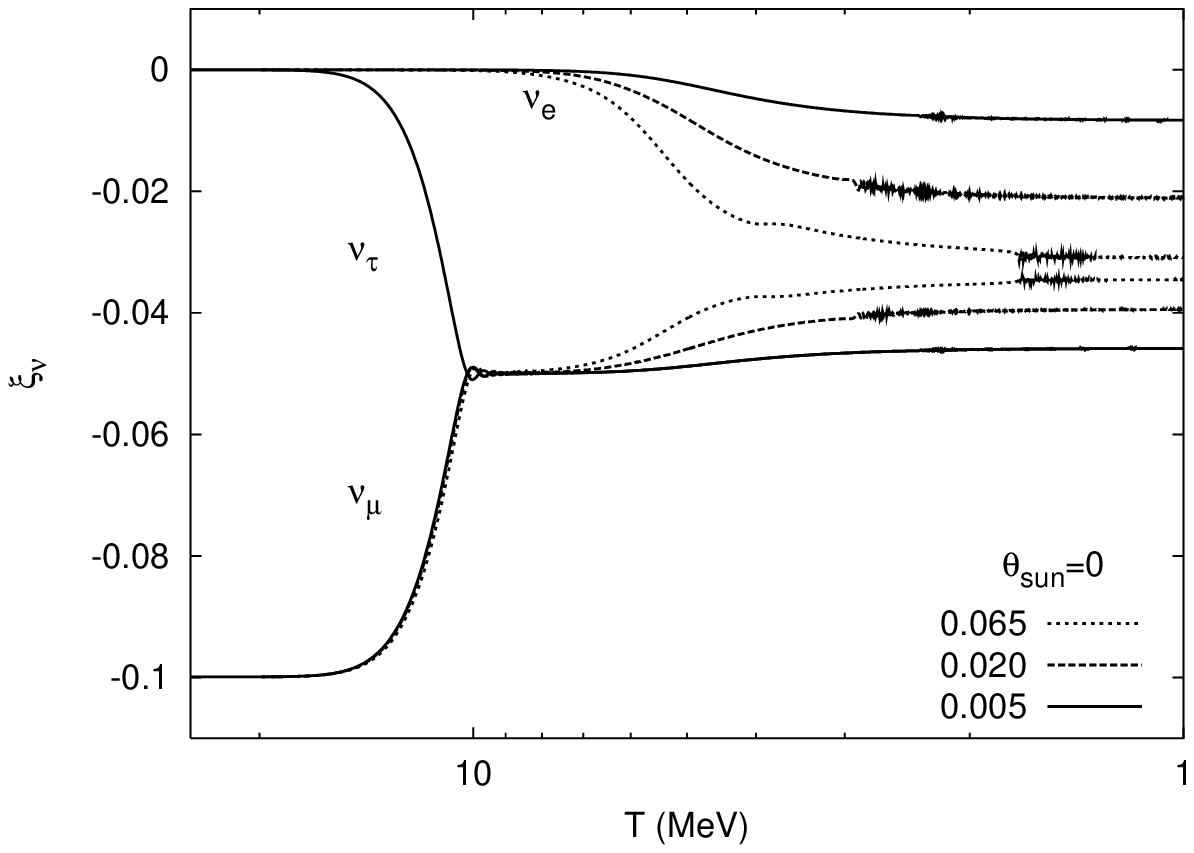,width=12cm}}
\caption{Evolution of the neutrino degeneracies for different values
of $\tan^2 \theta_{13}$ in the limit $\Delta m^2_{\rm sun} \simeq 0$
and $\theta_{\rm sun}\simeq 0$ for the initial conditions
$\xi_e=\xi_\tau=0$ and $\xi_\mu=-0.1$.}
\label{fig:8}
\end{figure}

The angle $\theta_{13}$ is restricted to the approximate region
$\tan^2 \theta_{13}\lsim 0.065$ (see for
instance~\cite{Bandyopadhyay:2001fb}) from a combined analysis of
solar, atmospheric and reactor (CHOOZ) data.  However, a small but
non-zero $\theta_{13}$ does modify the oscillation behavior. This
effect is shown in Fig.~\ref{fig:7} for the LMA case and different
values of $\theta_{13}$. Even small values of $\theta_{13}$ lead to
conversion to the electronic flavor at larger temperatures and enhance
flavor equilibration if $\theta_{\rm sun}$ is in the LMA region.

The effect of a non-zero $\theta_{13}$ is more important if the solar
neutrino problem is solved by oscillations with parameters in a region
other than LMA. This can be seen in Fig.~\ref{fig:8}, where we have
calculated the evolution in the limit $\Delta m^2_{\rm sun} \simeq 0$
and $\theta_{\rm sun} \simeq 0$, thus corresponding approximately to
the SMA, LOW and Vacuum regions. For values of $\tan^2 \theta_{13}$
close to the limit discussed at the beginning of this section,
neutrino oscillations lead to almost complete flavor equilibration,
while for smaller $\theta_{13}$ angles the conversion is only partial.

\subsection{Can the self-term prevent flavor equilibrium?}

In the two-flavor case of Sec.~\ref{sec:equalopposite} we saw that for
equal but opposite asymmetries of the two flavor distributions the
synchronized oscillation frequency was strictly zero, suppressing
oscillations entirely before the BBN epoch. In the three flavor case
the situation is more complicated so that again we need to raise the
question if there is a special configuration of initial neutrino
distributions which suppresses flavor oscillations by the neutrino
self-potential.

\begin{figure}[t]
\centerline{\epsfig{file=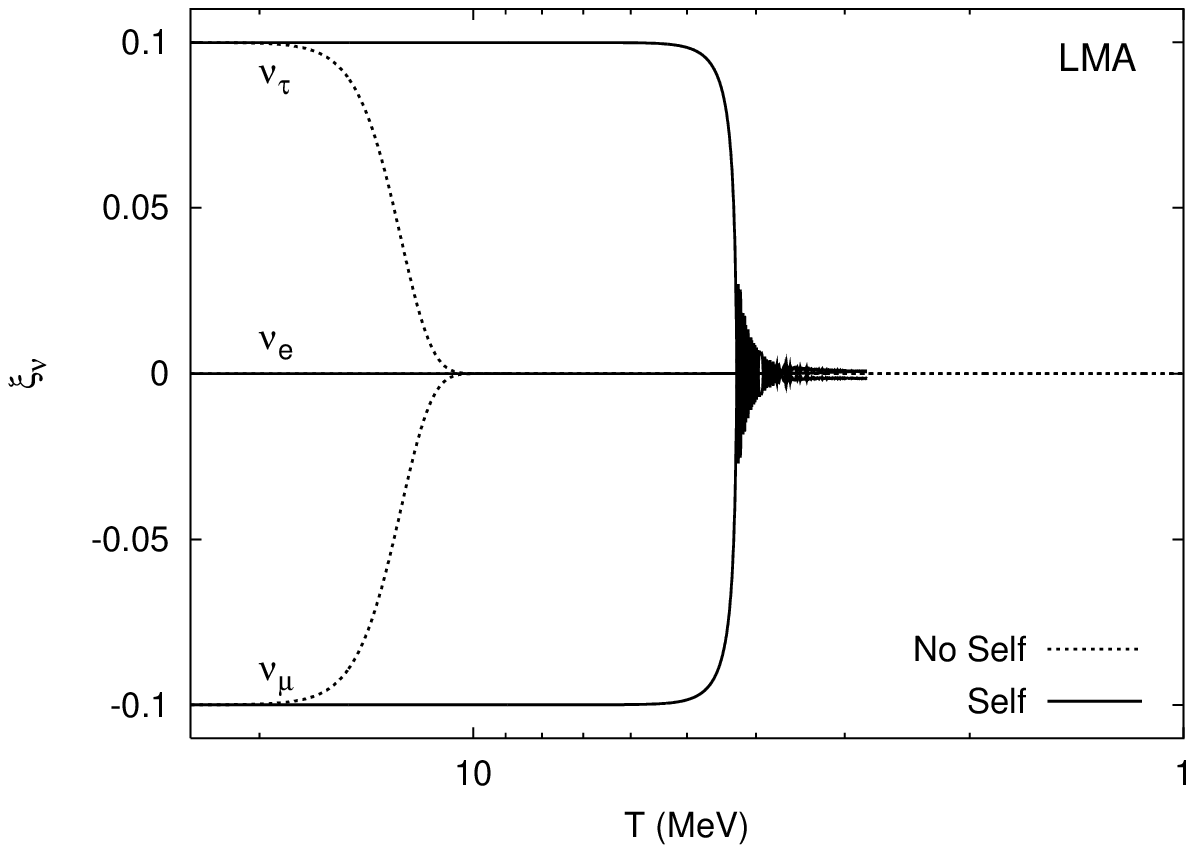,width=12cm}}
\caption{Evolution of the neutrino degeneracies if initially $\xi_e=0$
and $\xi_\tau=-\xi_\mu=0.1$, with or without neutrino self-interactions.}
\label{fig:9}
%
\bigskip
\centerline{\epsfig{file=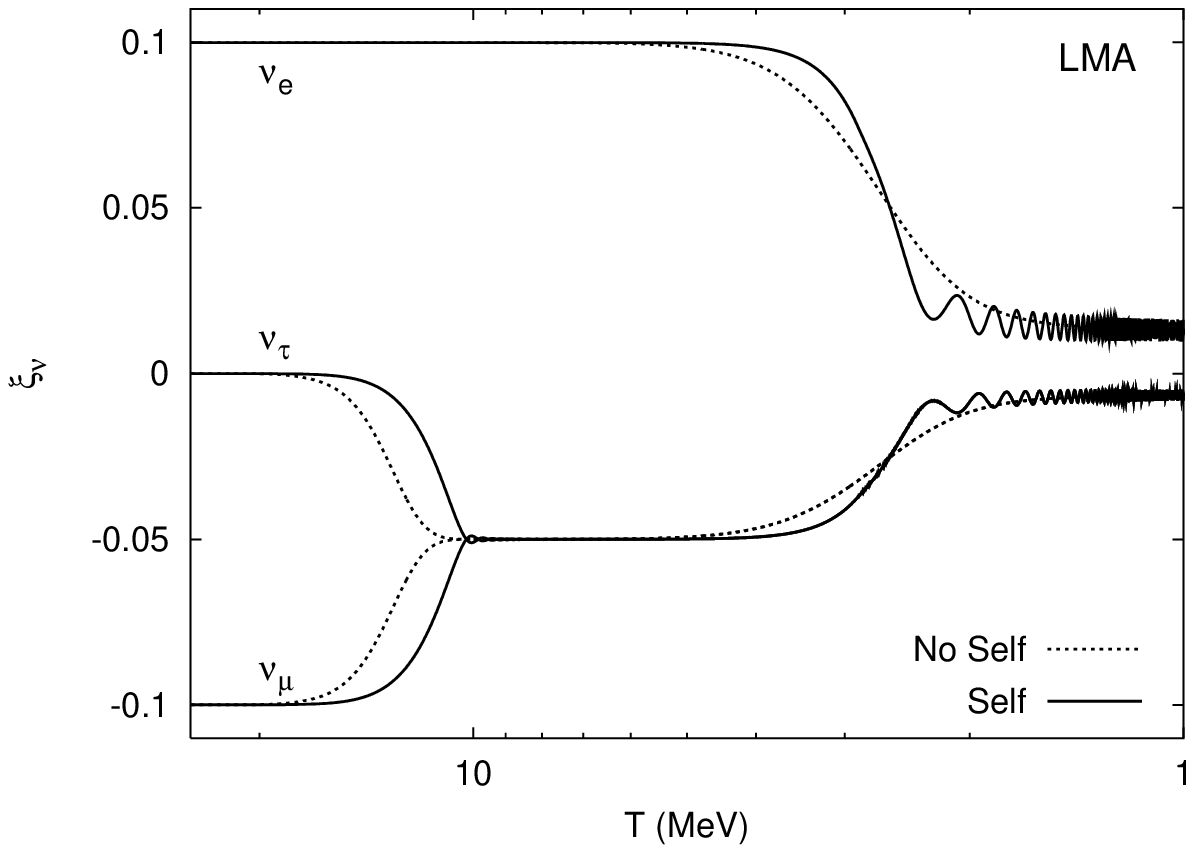,width=12cm}}
\caption{Evolution of the neutrino degeneracies for the LMA case with
$\theta_{13}=0$ with initial conditions $\xi_\tau=0$ and
$\xi_e=-\xi_\mu=0.1$, with or without neutrino self-interactions.}
\label{fig:10}
\end{figure}
In analogy to Sec.~\ref{sec:equalopposite} we first consider
situations where the asymmetries of two flavors are equal but
opposite, while the third asymmetry is strictly zero. In this case one
expects that oscillations between the asymmetric flavors are
suppressed by the self-potential while oscillations into the third
flavor are unimpeded. We show the numerical results for two specific
cases. In Fig.~\ref{fig:9} we take $\xi_\mu=-\xi_\tau=-0.1$ and
$\xi_e=0$.  The ``atmospheric'' oscillations between $\nu_\mu$ and
$\nu_\tau$ are indeed blocked in the presence of self-interactions
while the ``solar oscillations'', here the LMA case and $\tan^2
\theta_{13}=0$, proceed at the appropriate temperature.
In Fig.~\ref{fig:10} we take $\xi_\tau=0$ and $\xi_e=-\xi_\mu=0.1$.
Both the atmospheric and solar oscillations take place and equilibrate
the flavors essentially as in the previous cases with a simple initial
condition.

The question remains if one could devise special initial conditions
such that $|\xi_\tau|$ and $|\xi_\mu|$ are large while 
$|\xi_e|$ is small, and that this system avoids flavor oscillations by
reducing the synchronized oscillation frequency to a very small value.
In this case our new limits would not apply.

We do not believe that such a case can be constructed. To argue in
favor of this claim we first note that the most general initial
condition consists of thermal equilibrium distributions which are
diagonal in flavor space, i.e.\ in the weak-interaction basis the
initial $3{\times}3$ density matrices for neutrinos and anti-neutrinos
are diagonal and characterized by Fermi-Dirac distributions.
All off-diagonal elements would be quickly damped by reactions which
involve muons and electrons and thus ``measure'' the flavor content of
the participating neutrinos. Therefore, the most natural initial
condition for our problem is indeed characterized by $\xi_e$,
$\xi_\mu$ and $\xi_\tau$. 

The general equations of motion Eq.~(\ref{eq:3by3evol}) reveal that the
quantity which develops in a synchronized fashion is $\rho-\bar\rho$,
in agreement with the discussion of $2{\times}2$ synchronized
oscillations in Ref.~\cite{Pastor:2001iu} where the picture of
polarization vectors and the associated ``internal magnetic fields''
of the system was used. Therefore, the synchronized equation of motion
is obtained by subtracting the two lines of Eq.~(\ref{eq:3by3evol})
and integrating over all modes,
\begin{eqnarray}
i\partial_t(\rho-\bar\rho)&=&
\int\frac{d^3{p}}{(2\pi)^3}
\left[\frac{M^2}{2p},(\rho_p+\bar\rho_p)\right]\nonumber\\
&&{}-\sqrt2 G_{\rm F}\int\frac{d^3{p}}{(2\pi)^3}
\left[\frac{8p}{3 m_{\rm W}^2}{E},
(\rho_p+\bar\rho_p)\right]\nonumber\\
&&{}+\sqrt2 G_{\rm F}
\left[(\rho-\bar\rho),(\rho-\bar\rho)\right]~,
\end{eqnarray}
where we have dropped the collision term.  The last term, of course,
is identically zero. The second term vanishes also as long as the
density matrices are diagonal in the flavor basis since the matrix $E$
is diagonal in that basis. Only the first term, involving the mass
matrix, represents a ``force'' such as to move $\rho-\bar\rho$ away
from being diagonal in the flavor basis, i.e.\ which causes
synchronized flavor oscillations. This force is identically zero if
the matrix
\begin{equation}
\int\frac{d^3{p}}{(2\pi)^3}\,
\frac{\rho_p+\bar\rho_p}{2p}
\end{equation}
is proportional to the unit matrix. As the matrices $\rho_p$ and
$\bar\rho_p$ are initially 
diagonal and given by Fermi-Dirac distributions, the integral can be
solved explicitly, leading to an expression proportional to 
\begin{equation}
\pmatrix{1&&\cr&1&\cr&&1\cr}+
\frac{3}{\pi^2}
\pmatrix{\xi_e^2&&\cr&\xi_\mu^2&\cr&&\xi_\tau^2\cr}~.
\end{equation}
If all $\xi$ are initially equal, then the neutrinos are already in
flavor equilibrium and indeed oscillations trivially do not operate.

The only nontrivial configuration where synchronized oscillations
proceed with a vanishing frequency is when one of the chemical
potentials is opposite in sign to the others.  We have checked
numerically that indeed no flavor conversion arises in this case as
long as the self-potential is large, i.e.\ as long as we are in the
synchronized regime.  In the two-flavor case this corresponds to the
situation discussed in Sec.~\ref{sec:equalopposite}.

As in the two-flavor case we conclude that it is possible to avoid
flavor equilibrium by specially chosen initial conditions, but these
conditions require $|\xi_e|=|\xi_\mu|=|\xi_\tau|$. Therefore, the
strict BBN limits on $|\xi_e|$ apply to all flavors.


\section{New limits on neutrino degeneracy}

\label{sec:newlimits}

We conclude that in the LMA region the neutrino flavors essentially
equilibrate long before $n/p$ freeze out, even when $\theta_{13}$ is
vanishingly small.  For the other cases the outcome depends on the
magnitude of $\theta_{13}$.  In the LMA case it is thus justified to
derive new limits on the cosmic neutrino degeneracy parameters under
the assumption that all three neutrino flavors are characterized by a
single degeneracy parameter, independently of the primordial initial
conditions. We do not derive the corresponding limits for the other
solar neutrino solutions, since they would strongly depend on the
value of a non-zero $\theta_{13}$. However, if that angle is close to
the experimental limit, the bounds that we describe would
approximately apply.

We first note that the energy density in one species of neutrinos and
anti-neutrinos with degeneracy parameter $\xi$ is
\begin{equation}
\rho_{\nu\bar\nu}=T_\nu^4\,\frac{7\pi^2}{120}\left[1+
\frac{30}{7}\left(\frac{\xi}{\pi}\right)^2+
\frac{15}{7}\left(\frac{\xi}{\pi}\right)^4\right].
\end{equation}
It is clear that the BBN limit will imply $\xi\ll 1$ for all flavors
so that the modified energy density and the resulting change of the
primordial helium abundance $Y_{\rm p}$ will be negligibly
small. If there are additional relativistic species, such as sterile
neutrinos or majorons, then Eq.~(\ref{nooscbounds}) will simply apply 
to all the active neutrinos
\begin{equation}
\left| \xi \right| < 0.22 \, .
\end{equation}

Therefore, the only remaining BBN effect is the shift of the beta
equilibrium by $\xi_e$. We recall that $Y_{\rm p}$ is essentially
given by $n/p$ at the weak-interaction freeze-out, and that
$n/p\propto \exp(-\xi_e)\simeq 1-\xi_e$ where the latter expansion
applies for $|\xi_e|\ll 1$. Therefore, $\Delta Y_{\rm p}\simeq -Y_{\rm
p}(1-Y_{\rm p}/2)\xi_e \simeq -0.21\,\xi_e$. Modifications of $Y_{\rm
p}$ by new physics are frequently expressed in terms of the equivalent
number of neutrino flavors $\Delta N_\nu$ which would cause the same
modification due to the changed expansion rate at BBN. If the
radiation density at BBN is expressed in terms of $N_\nu$, the helium
yield can be expressed by the empirical formula $\Delta Y_{\rm
p}=0.012\,\Delta N_\nu$~\cite{Malaney:1993ah}. Therefore, the effect
of a small $\xi_e$ on the helium abundance is equivalent to $\Delta
N_\nu\simeq -18\xi_e$. A conservative standard limits holds that BBN
implies $|\Delta N_\nu|<1$ which thus translates into $|\xi_e|\lsim
0.057$. 

A more detailed recent analysis reveals that the measured primordial
helium abundance implies a 95\% CL range $N_\nu=2.5\pm0.8$ or $\Delta
N_\nu=-0.5\pm0.8$~\cite{Esposito:2000hh,Hansen:2001hi}.  We conclude
that the BBN-favored range for the electron neutrino degeneracy
parameter is at 95\% CL
\begin{equation}
\xi_e=0.03\pm0.04~.
\end{equation}
If all degeneracy parameters are the same, then this range applies to
all flavors.

It should be noted that the actual limit we obtain on
the neutrino degeneracy depends on the adopted BBN analysis. For
instance $\Delta N_\nu$ could be as high as $1.2$ when the primordial
abundance of lithium is used instead of that of deuterium
\cite{Olive:2002qg}. At any rate, a limit of $|\xi_e|\lsim 0.1$ seems
rather conservative and does not modify our conclusions.

Using $|\xi|<0.07$ as a limit on the one degeneracy parameter for all
flavors, the extra radiation density is limited by
$(\Delta\rho_{\nu\bar\nu})/\rho_{\nu\bar\nu}<3\times0.0021=0.0064$,
i.e.\ $\Delta N_\nu<0.0064$. If the same radiation density were to be
produced by the asymmetry of one single species, this would correspond
to $|\xi|<0.12$.  

For comparison with the future satellite experiments MAP and PLANCK
that will measure the CMBR anisotropies, it was calculated that they
optimistically will be sensitive to a single-species $\xi$ above $0.5$
and $0.25$, respectively~\cite{Kinney:1999pd}. However, with proper
consideration of the degeneracy with the matter density, $\omega_M$,
and the spectral index, $n$, a more realistic sensitivity is $\xi
\approx 2.4$ and $0.73$, respectively~\cite{Bowen:2001in}.  Turning
this around we conclude that our new limits are so restrictive that
the CMBR is certain to remain unaffected by neutrino degeneracy
effects so that $|\xi|$ can be safely neglected as a fit parameter in
future analyses.

If our new limits apply the number density of relic neutrinos is very
close to its standard value. Therefore, existing limits and possible
future measurements of the absolute neutrino mass scale, for example
in the forthcoming tritium decay experiment KATRIN \cite{KATRIN}, will
provide unambiguous information on the cosmic mass density in
neutrinos, free of the uncertainty of neutrino chemical potentials.

\noindent
{\bf Note added}: Very recently the authors of
refs.~\cite{Wong,Abazajianetal} have also analyzed the equilibration
of neutrino asymmetry from flavor oscillations, providing further
analytical insight and confirming our conclusions.

\section*{Acknowledgments}

We thank Alexei Smirnov, Gary Steigman, Karsten Jedamzik and Gianpiero
Mangano for useful discussions and comments. In Munich, this work was
partly supported by the Deut\-sche For\-schungs\-ge\-mein\-schaft
under grant No.\ SFB 375 and the ESF network Neutrino
Astrophysics. S.H.~Hansen and S.~Pastor are supported by Marie Curie
fellowships of the European Commission under contracts
HPMFCT-2000-00607 and HPMFCT-2000-00445.

\appendix
\section{Evolution equations}

In this appendix we list in detail the evolution equations for the
neutrino and anti-neutrino density matrices in the two-flavor case as
in Eq.~(\ref{2flavors}), while the generalization to the three-flavor
case that we consider in Sec.~\ref{sec:threeflavor} is
straightforward.

In our treatment of primordial neutrino oscillations we use the
following dimensionless expansion rate and momenta
\begin{equation}
x \equiv mR~, \qquad y \equiv pR~,
\end{equation}
where $R$ is the universe scale factor and $m$ an arbitrary mass scale
that we choose to be $1$ MeV. The neutrino and anti-neutrino density
matrices are
\begin{equation}
\rho(x,y) = 
\frac{1}{2} \left [P_0(x,y) +\hbox{\boldmath$\sigma$}
\cdot{\bf P}(x,y)\right]
=\frac{1}{2} \left (\matrix{P_0+P_z &
P_x-iP_y \cr P_x+iP_y&
P_0-P_z}\right )
\end{equation}
\begin{equation}
\overline{\rho}(x,y) = 
\frac{1}{2} \left [
\overline{P}_0(x,y)+\hbox{\boldmath$\sigma$}
\cdot{\bf \overline P}(x,y)\right]
=\frac{1}{2}  \left (\matrix{\overline{P}_0+\overline{P}_z &
\overline{P}_x-i\overline{P}_y \cr \overline{P}_x+i\overline{P}_y&
\overline{P}_0-\overline{P}_z}\right )
\end{equation}
with the derivatives
\begin{equation}
Hx \frac{d\rho}{dx}(x,y)=\frac{d\rho}{dt}(t,p)~,
\qquad
Hx \frac{d\overline{\rho}}{dx}(x,y)=
\frac{d\overline{\rho}}{dt}(t,p)~.
\end{equation}
The initial conditions are for large temperatures as follows. For
density matrices normalized to $f_{\rm eq}(y) = (e^y+1)^{-1}$ and
initial degeneracies $\xi_\alpha=-\xi_{\bar{\alpha}}$ and
$\xi_\beta=-\xi_{\bar{\beta}}$ (for flavor neutrinos $\nu_\alpha$ and
$\nu_\beta$)
\begin{eqnarray}
P_0(y)&=& 
\frac{f_{\rm eq}(y,\xi_\alpha)+f_{\rm eq}(y,\xi_\beta)}{f_{\rm eq}(y)}
\nonumber\\
P_z(y) &=& \frac{f_{\rm eq}(y,\xi_\alpha)-
f_{\rm eq}(y,\xi_\beta)}{f_{\rm eq}(y)}
\nonumber\\
\overline{P}_0(y) &=& 
\frac{f_{\rm eq}(y,-\xi_\alpha)+f_{\rm eq}(y,-\xi_\beta)}{f_{\rm eq}(y)}
\nonumber\\
\overline{P}_z(y) &=& 
\frac{f_{\rm eq}(y,-\xi_\alpha)-f_{\rm eq}(y,-\xi_\beta)}{f_{\rm eq}(y)}
\end{eqnarray}
where $f_{\rm eq}(y,\xi) = 1/[\exp(y-\xi)+1]$. Finally,
\begin{equation}
P_x(y) = P_y(y) = \overline{P}_x(y)= \overline{P}_y(y)=0~.
\end{equation}
The components of the neutrino polarization vectors evolve as
\begin{eqnarray}
\frac{dP_0}{dx} (x,y)&=& R_\alpha(x,y)+ R_\beta(x,y)~,\nonumber\\
\frac{dP_x}{dx} (x,y)&=&
\left(\frac{\Delta m^2}{2p Hx}~\cos 2\theta 
- \frac{V_l}{Hx} \right)~P_y(x,y)- \frac{D}{Hx}~P_x(x,y)~,\nonumber\\
\frac{dP_y}{dx} (x,y)&=&
-\left(\frac{\Delta m^2}{2p Hx}~\cos 2\theta 
- \frac{V_l}{Hx} \right)~P_x(x,y)
- \frac{D}{Hx}~P_y(x,y) \nonumber\\
&&- \frac{\Delta m^2}{2p Hx}~\sin 2\theta ~P_z(x,y)~,\nonumber\\
\frac{dP_z}{dx} (x,y)&=&
\frac{\Delta m^2}{2p Hx}~\sin 2\theta ~P_y(x,y)
+R_\alpha(x,y)- R_\beta(x,y)~,\nonumber\\
\frac{d\overline{P}_0}{dx} (x,y)&=& 
\overline{R}_\alpha(x,y)+ \overline{R}_\beta(x,y)~,\nonumber\\
\frac{d\overline{P}_x}{dx} (x,y)&=&
-\left(\frac{\Delta m^2}{2p Hx}~\cos 2\theta 
- \frac{V_l}{Hx} \right)~\overline{P}_y(x,y) 
- \frac{D}{Hx}~\overline{P}_x(x,y)~,\nonumber\\
\frac{d\overline{P}_y}{dx} (x,y)&=&
\left(\frac{\Delta m^2}{2p Hx}~\cos 2\theta 
- \frac{V_l}{Hx} \right)~\overline{P}_x(x,y)
- \frac{D}{Hx}~\overline{P}_y(x,y) \nonumber\\
&&+ \frac{\Delta m^2}{2p Hx}~\sin 2\theta ~\overline{P}_z(x,y)~,\nonumber\\
\frac{d\overline{P}_z}{dx} (x,y)&=&
-\frac{\Delta m^2}{2p Hx}~\sin 2\theta ~\overline{P}_y(x,y)
+\overline{R}_\alpha(x,y)- \overline{R}_\beta(x,y)~.
\label{dpidx}
\end{eqnarray}
where
\begin{eqnarray}
R_\alpha(x,y)&=& 2 \frac{D}{Hx} \left[
\frac{f_{\rm eq}(y,\xi_\alpha)}{f_{\rm eq}(y)}
-\frac{1}{2} \left (P_0(x,y)+P_z(x,y)\right)
\right]\nonumber\\
R_\beta(x,y)&=& 2 \frac{D}{Hx} \left[
\frac{f_{\rm eq}(y,\xi_\beta)}{f_{\rm eq}(y)}
-\frac{1}{2} \left (P_0(x,y)-P_z(x,y)\right)
\right]~,
\end{eqnarray}
with $\overline{R}_{\alpha,\beta}$ given by the same expressions with
$\xi \to -\xi$.  However, our numerical results do not change
significantly if the total neutrino and anti-neutrino number densities
are taken to be constant ($dP_0/dx=d\overline{P}_0/dx=0$).

The different terms in Eqs.~(\ref{dpidx})are given as follows. The
vacuum oscillation terms are proportional to
\begin{equation}
\frac{\Delta m^2}{2p Hx}=\frac{10^{10} ~M_P}{2\sqrt{8\pi/3}}~
\left(\frac{\Delta m^2}{\mbox{eV}^2}\right)
\frac{1}{\sqrt{\bar{\rho}}}~
\frac{x^2}{y}
\end{equation}
where $M_P \equiv 1.221$, $\bar{\rho}=(x/m)^4 \rho_{\rm tot}$, and
$\rho_{\rm tot}$ is the total energy density of the universe.
The $l^+l^-$ background with $l=\mu,e$ the charged leptons (for
$\nu_\tau$-$\nu_\mu$ and $\nu_e$-$\nu_{\mu,\tau}$ oscillations,
respectively) is described by
\begin{equation}
\frac{V_l}{Hx} = -\frac{8\sqrt{2}~10^5~M_P ~G_F}
{3~\sqrt{8\pi/3}~m_W^2}
~\frac{1}{\sqrt{\bar{\rho}}}~
\frac{y}{x^4}
~\left(\bar{\rho}_{l^+}+\bar{\rho}_{l^-}\right)
\label{VeHx}
\end{equation}
where $G_F \equiv 1.1664$ and $m_W \equiv 80.42$.
Collisions are described by damping terms 
where, according to our calculations,
\begin{eqnarray}
D &\simeq & 2 \times (4 \sin^4 \theta_W -
2 \sin^2 \theta_W + 2) F_0~,\nonumber\\
D &\simeq & 2 \times (2 \sin^4 \theta_W +
6) F_0~,
\end{eqnarray}
for $\nu_\tau$-$\nu_\mu$ and $\nu_e$-$\nu_{\mu,\tau}$ oscillations,
respectively, where all collision terms are proportional to
\begin{equation}
\frac{F_0}{Hx} = \frac{M_P ~G_F^2~\zeta(3)~y_{\rm med}}
{3~\pi^3~\sqrt{8\pi/3}}
~\frac{1}{\sqrt{\bar{\rho}}}~
\frac{y}{x^4}
\end{equation}
and $\zeta(3) \simeq 1.20206$ and $y_{\rm med} \simeq 3.15137$.

Finally one has to add the contribution from the neutrino-antineutrino
background \cite{Sigl:1993fn,McKellar:1994ja,Pantaleone:1992eq}
\begin{eqnarray}
\left[
\frac{V_{\rm asym}}{Hx}~\left({\bf J}(x)-{\bf \overline J}(x)\right)
- \frac{V_{\rm sym}}{Hx}~\left({\bf U}(x)+{\bf \overline U}(x)\right)
\right]
\times {\bf P}(x,y)~,\nonumber\\
\left[
\frac{V_{\rm asym}}{Hx}~\left({\bf J}(x)-{\bf \overline J}(x)\right)
+ \frac{V_{\rm sym}}{Hx}~\left({\bf U}(x)+{\bf \overline U}(x)\right)
\right]
\times {\bf \overline P}(x,y)~,
\end{eqnarray}
for $d{\bf P}/dx$ and $d{\bf \overline P}/dx$, respectively. Here we
have defined
\begin{equation}
{\bf J}(x) = \frac{1}{2\pi^2}\int du~u^2~{\bf P}(x,u)~,
\qquad
{\bf U}(x) = \frac{1}{2\pi^2}\int du~u^3~{\bf P}(x,u)~,
\end{equation}
and the same for ${\bf \overline J}$ and ${\bf \overline U}$. Moreover,
\begin{eqnarray}
\frac{V_{\rm asym}}{Hx}&=& 
\frac{\sqrt{2}~10^{11}~M_P ~G_F}
{\sqrt{8\pi/3}}
~\frac{1}{\sqrt{\bar{\rho}}~x^2}\nonumber\\
\frac{V_{\rm sym}}{Hx}&=& 
\frac{8\sqrt{2}~10^5~M_P ~G_F}
{3~\sqrt{8\pi/3}~m_W^2}~\cos^2 \theta_W~
~\frac{1}{\sqrt{\bar{\rho}}}~
\frac{y}{x^4}~.
\end{eqnarray}

\end{document}